\documentclass{elsart}
 \journal{Applied Mathematics and
     Computation}
 \usepackage{graphicx}
 \usepackage{amssymb}
 \usepackage{fancyhdr}
 \usepackage{rotating}

\newcommand{\be}{\begin{equation}}
\newcommand{\en}{\end{equation}}

\newcommand{\ee}{\end{equation}}

\newcommand{\ba}{\begin{eqnarray}}
\newcommand{\ea}{\end{eqnarray}}

\newcommand{\bea}{\begin{eqnarray}}
\newcommand{\ena}{\end{eqnarray}}

\newcommand{\beano}{\begin{eqnarray*}}
\newcommand{\enano}{\end{eqnarray*}}

\newcommand{\bei}{\begin{itemize}}
\newcommand{\eni}{\end{itemize}}

\newcommand{\bee}{\begin{enumerate}}
\newcommand{\ene}{\end{enumerate}}

\newtheorem{theorem}{Theorem}[section]
\newtheorem{coroll}[theorem]{Corollary}
\newtheorem{lemma}[theorem]{Lemma}

\newcommand{\betheo}{\begin{theorem}}
\newcommand{\enth}{\end{theorem}}

\newcommand{\becor}{\begin{coroll}}
\newcommand{\encor}{\end{coroll}}

\newcommand{\belem}{\begin{lemma}}
\newcommand{\enlem}{\end{lemma}}

\newcommand{\beprop}{\begin{prop}}
\newcommand{\enprop}{\end{prop}}

\begin{document}

\begin{frontmatter}
\title{Generalized Heun and Lam\'e's equations: factorization}
\author[cotonou]{Mahouton Norbert Hounkonnou\thanksref{X}}and
\thanks[X]{Correspondence should be addressed to:
 norbert.hounkonnou@cipma.uac.bj, with copy to hounkonnou@yahoo.fr}
% \ead{norbert$_{-}$hounkonnou@cipma.net}
\author[Louvain-la-Neuve]{Andr\'e Ronveaux}
\address[cotonou]{International Chair in Mathematical Physics
and Applications, \\(ICMPA - UNESCO Chair) 072 B.P.:50, Universit\'e
d'Abomey-Calavi, Cotonou, Republic of Benin}
\address[Louvain-la-Neuve] {D\'epartement de Math\'ematiques,
 Universit\'e catholique  de Louvain,
2 chemin du cyclotron, $B$-1348 Louvain-la-Neuve, Belgium}

\begin{abstract}
 This paper addresses new results on the factorization of the general
 Heun's operator, extending  the investigations performed in previous works
 [{\it Applied
 Mathematics and Computation} {\bf 141} (2003), 177 - 184 and {\bf 189} (2007), 816 - 820].
Both generalized Heun and generalized Lam\'e equations are
considered.
\end{abstract}
\begin{keyword}
 Factorization,  Heun's differential
equation,  Lam\'e's differential equation, polynomials.
\end{keyword}
\end{frontmatter}
ICMPA-MPA/2009/14
  \section{Introduction}
 Fuchsian differential's equations of second order and their confluent forms,
 built for instance from the Ince  techniques \cite{bi1} play a major role in many partial
 differential equations of mathematical physics as Laplace, Helmohltz or
 Schr\"{o}dinger's equations.

 Without loss of generality the $k+3$ regular single singularities can be located at
 $x=0$, $x=1$, $x=a_{i},\,\,i=\overline{1,\,k}$ and $x=\infty$ and after an appropriate change of function
 one of the indices at each finite singulary can be shifted to zero allowing
 to write the Fuchsian equation  as \cite{bi2,pool}:
 \begin{eqnarray}
 \label{eq1}
 y''(x)+
 \left(
 \frac{\gamma}{x}+\frac{\delta}{x-1} + \sum_{i=1}^{k}\frac{\epsilon_{i}}{x-a_{i}}
 \right)y'(x)
 %\nonumber\\
 +\frac{\alpha\beta x^{k}+ \sum_{i=1}^k\rho_ix^{k-i}
 }{x(x-1)\prod_{i=1}^{k}(x-a_{i})}y(x)=0.\nonumber\\
 \end{eqnarray}
 The two indices at each singularity $a_{i}$ are $(0, 1-\epsilon_{i})$
 and $(0,1-\gamma)$, $(0,1-\delta)$,   $(\alpha,\beta)$ at
 $x=0,$ $x=1$, and $x=\infty$, respectively, taking into account the Fuchsian relation:
 \begin{equation}\label{eq2}
 \alpha+\beta+1 = \gamma+\delta+\sum_{i=1}^{k}\epsilon_{i}.
 \end{equation}
 $\epsilon_{i}\equiv 0, \, (i=\overline{1,\,k})$, gives the hypergeometric  equation and $k=1$ the
 Heun's equation \cite{bi6}. The character {\it reducible} or {\it irreducible} of linear
 ordinary differential equation (O.D.E)  is important  in relation with the monodromy
 group \cite{bi2}. We call reducible a second order O.D.E which admits a non trivial
 solution satisfying  a first order linear equation, and otherwise irreducible. This definition
 is used by many authors \cite{bi3,bi4,bi5} with or without supplementary conditions.
 It  allows to give an easy criterion of reducibility of an equation with polynomial coefficients,
  from a factorized  approach in the following way.

Rewriting (\ref{eq1}) in a polynomial form with $D=\frac{\rm d}{{\rm d}x}$,
\begin{eqnarray}\label{eq3}
\label{eq3} {\cal H}_{k}[y(x)]&\equiv&\left[Q_{k+2}(x)D^{2}+Q_{k+1}D
+Q_{k}\right]y(x)\cr &=& \left[L(x)D + M(x)\right]\left[\bar{L}(x)D
+ \bar{M}(x)\right]y(x),
\end{eqnarray}
where $Q_j$ are polynomials of degree $j,$ possible (or not
possible) identification of the polynomials $L$, $\bar{L}$, $M$,
$\bar{M}$ with polynomials $Q_{k+2}$, $Q_{k+1}$, $Q_{k}$ will give
all possible cases  of reducibility and, a contrario, criteria of
irreducibility for the equation ${\cal H}_{k}[y(x)]=0$.

This was recently done   for the Heun's equation \cite{bi7,bi8}
\begin{equation}
\label{eq4} y''(x)+\left[
\frac{\gamma}{x}+\frac{\delta}{x-1}+\frac{\epsilon}{x-a}
\right]y'(x) +\frac{\alpha\beta x-q}{x(x-1)(x-a)}y(x)=0
\end{equation}
and also for the  four confluent Heun's equations \cite{bi9}. For
the Heun's equation, this factorization is possible and generate 6
non trivial
 situations in accordance with the known $F-$homotopic transformations \cite{bi10}.
R. S Maier \cite{bi11}  noticed that for the Lam\'e
  equation $(k=1 \; \mbox{with}\;\gamma= \delta= \epsilon= 1/2)$
 \begin{equation}
 \label{eq5}
 y''(x)+\frac{1}{2}\left(
 \frac{1}{x}+\frac{1}{x-1}+\frac{1}{x-a}\right)y'(x)-\frac{l(l+1)x+4q}
 {4x(x-1)(x-a)}y(x)=0,
 \end{equation}
 the factorization reduces only to the two cases $l=1$,
 $l=\frac{1}{2}$. He
   also  mentioned that for the general Lam\'e equation with $k+3$
  singularities, the factorization is not known.

  Even if  some algorithms exist in order to factorize any ODE with
  rational function coefficients
   \cite{bi12,bi13,bi14,bi15}, factorization of an arbitrary
   $k-$Heun equation is probably untractable, except for using  numerical tools.

   The aim of this work is to investigate the case $k=2$  for both generalized
    Heun and generalized Lam\'e equations \cite{bi15a}
   which are also important in lattice statistics \cite{bi16}.
These equations have invested a large number of applications in
physics. See \cite {biphys} (and references therein) for a nice
review on physical applications of these equations.   To mention a
few, retrieved from the indicated works, where for instance Heun's
equations as well as their
   solutions impose their usefulness, it is worthy of attention to outline their
   importance in the description  of
   quasi-modes of near extremal black branes,
   in the hyperspherical harmonics with applications in three-body systems,
    in the elaboration of a method of calculation of propagators for the
    case of a massive spin 3/2 field for arbitrary space-time dimensions
    and mass, in parametric resonance after inflation, as well as in the
     separation of variables for the Schr$\ddot o$dinger equation in a large
      number of problems, typically for the radial coordinate, and in
       non-linear formulations involving Painlev´e type equations.
       A number of traditional equations of mathematical physics, as for instance the Lam´e,
   spheroidal wave, and Mathieu equations, are also particular cases of Heun equations.

Expansion of the equation (\ref{eq3}) gives:
\begin{eqnarray}\label{eq6}
{\cal H}_{k}[y(x)]&=& (LD+M)\left(\bar L D+\bar M\right)y(x)\nonumber\\
&=&\left\{L\bar LD^2 +\left[L\left(\bar L^\prime+\bar M\right) +
M\bar L\right] D + \left(L\bar
M^\prime + M\bar M\right)\right\}y(x)\nonumber\\
\end{eqnarray}
and generate the 3 basic relations:
\begin{equation}\label{eqn0}
\left\{\begin{array}{lll}
 Q_{k+2}& =& L\bar L,\\
 Q_{k+1} &=& L\left(\bar L^\prime+\bar
M\right) + M\bar L,\\
Q_{k} &=& L\bar M^\prime +M\bar M.
\end{array}\right.
\end{equation}
Computation and properties of the polynomials $M\equiv M(x), \;\bar
M\equiv \bar M(x), \;L\equiv L(x), \;\bar L\equiv \bar L(x)$ are
explicitly given in several situations.
\section{Preliminary remarks}\label{sect2}
\begin{itemize}
\item[1)] $Q_{k+2}(x)$ contains $k+2$ linear and different factors which
are all present in the product $L(x)\bar L(x)$ in several ways. For
instance, if $j$ factors appear in $L$ and $k+2-j$ in $\bar L$, the
number of decomposition is given by the classical binomial result:
\begin{eqnarray}\label{eq10}
\sum_{j=0}^{k+2}\left(_j^{k+2}\right)=2^{k+2},
\end{eqnarray}
excluding the extremal cases
%\begin{itemize}
%\item
\begin{eqnarray}
L(x)&=& Q_{k+2}(x), \,\bar L=1;\;Q_k=0,\nonumber\\
{\cal H}_{k}&=&LD(D+\bar M) \label{eq11}
\end{eqnarray}
and
%\item
\begin{eqnarray}
 L(x)&=& 1, \, \bar L= Q_{k+2}(x),\, M=0 \;
(\mbox{integrable case}\,\bar M= Q_{k+1}-Q'_{k+2},\nonumber\\
Q_k&=&\bar M'),\;
 {\cal H}_{k}=D(\bar L D+\bar
M)\label{eq12}
\end{eqnarray}
%\end{itemize}
which are irrelevant or trivial in this study.
% From incompatibility of degree
%of the polynomials, if $Q_k\neq 0$ then, in the situation
%(\ref{eq11}), the system (\ref{eqn0}) does not have a solution while
%if $Q_k=0$ in the same situation the factorization takes the form
 %\begin{equation}
 %(L(x)D+M(x))\bar{L}(x)D\,y(x)=0.
%\end{equation}
%Then, in the case (\ref{eq12}), solution  of the system (\ref{eqn0})
%does not always exist. When it exists, factorization takes the form
%\begin{equation}
%L(x)D\,\left[\bar{L}(x)D+\bar{M}(x)\right]\,y(x)=0.
%\end{equation}

Selecting a pair $L, \, \bar L$ inside the $ 2^{k+2}-2$
possibilities, computation of the polynomials $M(x)$ and $\bar M(x)$
of degrees $m$ and $\bar m$, respectively, is done in 2 steps.
\begin{itemize}
\item [i)] The coefficients in $Q_{k+1}(x)$ and $Q_{k+2}(x)$ allow to linearly
compute the coefficients of $M(x)$ and $\bar M(x)$, of number
$m+\bar m\le k$ from the second and third relations in (\ref{eqn0}).
\item [ii)] The third relation in (\ref{eqn0}) gives the value of the
accessory parameters $\rho_i,\, (i={\overline{1,k}})$ and the
product $\alpha\beta$, using the polynomials $M$ and $\bar M$
computed in the first step.
\end{itemize}

\item[2)] The general $k-$Lam\'e's equation is a $k-$ Heun's equation with
 $\gamma= \delta= \epsilon= 1/2, \, (i= \overline 1,k)$. The  indices at each
finite singularity are equal to $(0,\,1/2)$ and the Fuchsian
relation (\ref{eq2}) gives $\alpha + \beta = k/2$. This equation
takes the form
\begin{eqnarray}\label{eq13}
{\cal L}_k[y]\equiv Q_{k+2}y'' +Q_{k+1}y' +{1 \over
4}\left(\alpha\beta x^k + \sum_{i=1}^{k}\rho_ix^{k-i}\right)y=0
\end{eqnarray}
with now
\begin{eqnarray}
Q_{k+1}&=&{1 \over 2}( Q_{k+2})^\prime= {1\over 2}\left[L\bar
L\right]^\prime\label{eq14},
\end{eqnarray}
where
\begin{eqnarray}
 Q_{k+2}&=&x(x-1)\prod_{j=1}^k(x-a_j)=L\bar L.\label{eq15}
\end{eqnarray}
The property (\ref{eq14}) and second  relation in (\ref{eqn0})
easily give:
\begin{eqnarray}\label{eq16}
{1\over 2}\left( {L\over {\bar L}}\right)^\prime = {L{\bar M
}+M{\bar L}\over {\bar L^2}}.
\end{eqnarray}
The number of situations to be investigated in the decomposition of
$L\bar L$ can be simplified from some symmetry between polynomials
$M,\,\bar M$ and $M^\star,\,\bar M^\star$ corresponding to the
decomposition $L\bar L$, permuting $L$ and $\bar L$ giving the
factorization:
\begin{eqnarray}\label{eq17}
{\cal L}_k[y]=\left[\bar LD+M^\star\right]\left[LD+\bar
M^\star\right]y(x).
\end{eqnarray}
A similar relation to (\ref{eq16}) is
\begin{eqnarray}\label{eq18}
{1\over 2}\left( {{\bar L}\over { L}}\right)^\prime = {\bar L{\bar
M^\star }+M^\star{L}\over {L^2}}.
\end{eqnarray}
  We deduce from (\ref{eq16})
and (\ref{eq18}) that
\begin{eqnarray}\label{eq19}
L\left(\bar M+M^\star\right)+\bar L\left( M + \bar M^\star\right)=
0.
\end{eqnarray}
Equality (\ref{eq19}) is satisfied if, for example, we have  $ M=
-\bar M^\star,\,\bar M=-M^\star$.
\end{itemize}

\section{Lam\'e's equation, ($k=1$)}
Let us consider equation (\ref{eq4}) with
$\gamma=\delta=\epsilon=\frac{1}{2}$. Even if this case has been
already solved in \cite{bi6}, as a subcase of the full Heun
factorization, we easily recover the polynomials $M$ and $\bar M$
from the procedure described in section \ref{sect2} for the
factorization $L= x(x-1), \, \bar L= x-a$ as example. Unknown
polynomials $M$ and $\bar M$ must be as:
\begin{eqnarray}\label{eq20}
M= Ax + B, \, \bar M= \bar A
\end{eqnarray}
with
\begin{eqnarray}
Q_3(x)&=& L\bar L= x^3-x^2(a+1)+x a ,\label{eq21}\\
Q_2(x)&=&L(\bar L^\prime + \bar M) + M\bar L= {3\over 2}x^2-x(a+1)+
{a\over 2}\label{eq22}\\
Q_1(x)&=&M\bar M.\label{eq23}
\end{eqnarray}
We easily conclude that
\begin{eqnarray}\label{eq23}
A= 1,\,B= -{1\over 2},\,\bar A= -{1\over 2}.
\end{eqnarray}
Next the third relation in (\ref{eqn0}) gives:
\begin{eqnarray}\label{eq23}
\alpha\beta= -{1\over 2},\,q= -{1\over 4}
\end{eqnarray}
and the Fuschian relation yields
\begin{eqnarray}\label{eq24}
\alpha + \beta= {1\over 2}.
\end{eqnarray}
From section 2, $M^\star= -\bar M$ and $\bar M^\star= - M$ already
solve the factorization problem with $L= x-a, \, \bar L= x(x-1)$.

A surprising result, consequence of a remark of Maier \cite{bi11},
is that for $L= x, \,\mbox{or}\, (x-1),\, \mbox{or}\,
(x-a),\,\left(\bar L= (x-1)(x-a),\,x(x-a),\,x(x-1)\right)$, the
polynomials $M$ are the same. From the symmetry property $M^\star=
-\bar M,\,\bar M^\star= - M$, this result also works for the $3$
other factorizations: $L= (x-1)(x-a),\,x(x-a),\,x(x-1);\,\bar L=
x,\,(x-1),\,(x-a)$. All these results can be seen in table 1,
presented in the same way  as in Ref \cite{bi8}.

 \begin{sidewaystable}[htbp]
%\begin{table}[htbp]
{\footnotesize
\begin{center}
\caption{Factorization of the simple Heun's  operator ${\cal
H}_1={\cal L}_1[y]$}
\begin{eqnarray}
{\cal H}_1&=&x(x-1)(x-a)D^2+\left[ \gamma(x-1)(x-a)+\delta
x(x-a)+\epsilon x(x-1)\right]D+(\alpha\beta x-q)I_d\cr
&=&(LD+M)(\bar L D+\bar M),\,\gamma=\delta= \epsilon= {1\over 2}.
\end{eqnarray}
%}
\vspace{.5cm}
\begin{tabular}{||l|l|l|l|l|l|l||}
\hline \hline
%& & & & & & &\\
 $L$ & ${\bar L}$ & $M$ & ${\overline M}$ & $\alpha$ & $\beta$ & $q$\\
\hline\hline
%& & & & & & & \\
 $x$  & $(x-1)(x-a)$ & ${1\over 2}$ &
 $-x +{{a+1}\over 2}$ & $ {3\over 2}$
  & $ -1 $ &$-{1\over 2}\left({(a+1)\over 2}\right)$ \\
\hline
%& & & & & & &\\
$x-1$  &$x(x-a)$ & $ {1\over 2} $& $-x +{a\over 2}$ &
  ${3\over 2}$
  &  $-1$ &$-\left(1+{a\over 4}\right)$\\
  \hline
%& & & & & & &\\
$x-a$  &$x(x-1)$ & $ {1\over 2} $& $-x +{1\over 2}$ &
  ${3\over 2}$
  &  $-1$ &$-\left(a+{1\over 4}\right)$\\
\hline
%& & & & & & &\\
$(x-1)(x-a)$  &$x$ & $ x-{{a+1}\over 2} $& $-{1\over 2}$ &
  $-{1\over 2}$
  &  $1$ &$-{1\over 4}(a+1)$\\
\hline
%& & & & & & &\\
$x(x-a)$  &$x-1$ & $ x-{{a}\over 2} $& $-{1\over 2}$ &
  $-{1\over 2}$
  &  $1$ &$-{1\over 4}a$\\
\hline
%& & & & & & &\\
$x(x-1)$  &$x-a$ & $ x-{1\over 2} $& $-{1\over 2}$ &
  $-{1\over 2}$
  &  $1$ &$-{1\over 4}$
\\
\hline \hline
\end{tabular}
\end{center}
} \label{tab1}
%\end{table}
 \end{sidewaystable}

\section{Factorization of Lame's equation ${\cal L}_2[y], \, k=2$}
Equation (\ref{eq1})  with $k= 2$, $a_1= a,\, a_2= b$ and
$\gamma=\delta=\epsilon_1=\epsilon_2=\frac{1}{2}$ is the Lam\'e
equation
\begin{eqnarray}\label{eq25}
{\cal L}_2[y]= Q_4(x)y'' + Q_3(x) y' + Q_2(x)y= 0,
\end{eqnarray}
where
\begin{equation}\label{eqn1}
\left\{\begin{array}{lll}
 Q_4(x)&=&x(x-1)(x-a)(x-b)\nonumber\\&=&x^4-x^3(1+a+b) +
 x^2(a+b+ab)-a b x,\label{eq26}\\
 Q_3(x)&=&{1\over 2}Q'_4(x)={1\over 2}\left[4x^3-3x^2(1+a+b)+2x(a+b+ab)-ab\right],\label{eq27}\\
Q_2(x)&=& \alpha\beta x^2+\rho_1x+\rho_2,\\
\alpha +\beta&=&1.\label{eq28}
\end{array}\right.
\end{equation}
With a first choice $L= x(x-1),\, \bar L=(x-a)(x-b)$, $M$ and $\bar
M$ are first degree polynomials
\begin{eqnarray}
M= Ax + B, \,\bar M= \bar Ax + \bar B.
\end{eqnarray}
The four equations satisfied by $A,\,B,\, \bar A,\,\bar B$ are given
by identification of second relation in (\ref{eqn0}) with the second
relation in (\ref{eqn1}). The solutions are
\begin{eqnarray}
A= 1,\, B=-\frac{1}{2},\,\bar A=-1,\,\bar B= \frac{a+b}{2}.
\end{eqnarray}
Comparison of third relation in (\ref{eqn0}) with the third relation
in  (\ref{eqn1}) yields
\begin{equation}
\alpha \beta=-2
\end{equation}
and also $\rho_1$ and $\rho_2$:
\begin{eqnarray}
\rho_1&=&\frac{a+b+3}{2},\\ \rho_2&=&-\frac{a+b}{4}.
\end{eqnarray}
%A second choice of the same type: $L=x(x-a),\,\bar L=(x-1)(x-b)$ or
%with asymmetric degree $L=x,\,\bar L=(x-1)(x-a)(x-b)$ gives

These results as well as those corresponding  to the other choices
of $L$ and $\bar L$ are reproduced in tables $2$ to $4$  ($14$
situations) for the cases
$\gamma=\delta=\epsilon_1=\epsilon_2=\frac{1}{2}$ as in the case $k=
1$, in which appear essentially $2$ situations \cite{bi11}; the case
$k= 2$ generates only $3$ situations with
$\alpha\beta=-2,\;-3/4,\;-15/4$.
\begin{sidewaystable}[h]
%\begin{table}[htbp]
{\footnotesize
\begin{center}
\caption{Factorization of the Lame's operator ${\cal L}_2[y]$ }
\begin{eqnarray*}
{\cal L}_2[y]&=&Q_4(x)y''+Q_3(x) y' +(\alpha\beta
x^2+\rho_1x+\rho_2)y= 0= (LD+M)(\bar L D+\bar M)y.\quad M= x+B, \,
\, \bar M=- x+\bar B.
\end{eqnarray*}

%}
\vspace{.5cm}
\begin{tabular}{||l|l|l|l|l|l|l|l||}
\hline \hline
%& & & & & & & &  \\
 $L $& ${\bar L}$ & $B$ & $\bar B$  &$\alpha$ & $\beta$&$\rho_1$& $\rho_2$ \\
\hline\hline
%& & & & & & & & \\
 $x(x-1)$&$(x-a)(x-b)$&
$-\frac{1}{2}$&$\frac{1}{2}(a+b)$&$-1$&$2$&$\frac{1}{2}(a+b+3)$&$-\frac{1}{4}(a+b)$\\
  \hline
$(x-a)(x-b)$&$x(x-1)$&$-\frac{1}{2}(a+b)$&$\frac{1}{2}$&$2$&$-1$&
$\frac{1}{2}(3a+3b+1)$&$-\frac{1}{4}(a+b+ab)$\\
\hline
$x(x-a)$&$(x-1)(x-b)$&$-\frac{1}{2}a$&$\frac{1}{2}(b+1)$&$-1$&$2$&$\frac{1}{2}(3a+b+1)$&$-\frac{1}{4}(a+ab)$\\
\hline
$(x-1)(x-b)$&$x(x-a)$&$-\frac{1}{2}(b+1)$&$\frac{1}{2}a$&$2$&$-1$&$\frac{1}{2}(a+3b+3)$&$-\frac{1}{4}(a+4b+ab)$\\
\hline
$x(x-b)$&$(x-1)(x-a)$&$-\frac{1}{2}b$&$\frac{1}{2}(a+1)$&$-1$&$2$&$\frac{1}{2}(a+3b+1)$&$-\frac{1}{4}(b+ab)$\\
\hline
$(x-1)(x-a)$&$x(x-b)$&$-\frac{1}{2}(a+1)$&$\frac{1}{2}b$&$2$&$-1$&$\frac{1}{2}(3a+b+3)$&$-\frac{1}{4}(4a+b+ab)$\\
\hline\hline
\end{tabular}
\end{center}
}\label{tab01}
%\end{table}
 \end{sidewaystable}

%%%%%%%%%%%%%%%%%%%%%%%%%%%%%%%%%%%%%%%%%%%%%%%%%%%%%%%%%%%%%%%%%%%%%%%%%%%%%%%%%%%%%%%%%%
%%%%%%%%%%%%%%%%%%%%%%%%%%%%%%%%%%%%%%%%%%%%%%%%%%%%%%%%%%%%%%%%%%%%%%%%%%%%%%%%%%%%%%%%%%

%\newpage
 \begin{sidewaystable}
%\begin{table}[htbp]
{\footnotesize
\begin{center}
\caption{Factorization of the Lame's operator ${\cal L}_2[y]$
(continuation). }
\begin{eqnarray*}
{\cal L}_2[y]&=&Q_4(x)y''+Q_3(x) y' +(\alpha\beta
x^2+\rho_1x+\rho_2)y= 0= (LD+M)(\bar L D+\bar M)y. \quad
M=\frac{3}{2} x^2+B x+C,\,\,\bar M =
-\frac{1}{2},\,\,\alpha=\frac{3}{2},\,\,\beta=-\frac{1}{2}.
\end{eqnarray*}
%}
\vspace{.5cm}
\begin{tabular}{||l||l|l|l|l|l||}
\hline \hline

$ L$ & ${\bar L}$ & $B$ & $C$ &$\rho_1$&$\rho_2$\\
\hline\hline
$(x-1)(x-a)(x-b)$&$x$&$-(a+b+1)$&$\frac{1}{2}(a+b+ab)$&$\frac{1}{2}(a+b+1)$&$-\frac{1}{4}(a+b+ab)$\\
\hline $x(x-a)(x-b)$&$x-1$&$-(a+b)$&$\frac{1}{2}ab$&$\frac{1}{2}(a+b)$&$-\frac{1}{4}ab$\\
\hline $x(x-1)(x-b)$&$x-a$&$-(b+1)$&$\frac{1}{2}b$&$\frac{1}{2}(b+1)$&$-\frac{1}{4}b$\\
\hline $x(x-a)(x-1)$&$x-b$&$-(a+1)$&$\frac{1}{2}a$&$\frac{1}{2}(a+1)$&$-\frac{1}{4}a$\\
 \hline \hline
\end{tabular}
\end{center}
} \label{tab02}
%\end{table}
 \end{sidewaystable}

%%%%%%%%%%%%%%%%%%%%%%%%%%%%%%%%%%%%%%%%%%%%%%%%%%%%%%%%%%%%%%%%%%%%%%%%
%%%%%%%%%%%%%%%%%%%%%%%%%%%%%%%%%%%%%%%%%%%%%%%%%%%%%%%%%%%%%%%%%%%%%%%%

%\newpage
 \begin{sidewaystable}
%\begin{table}[htbp]
{\footnotesize
\begin{center}
\caption{Factorization of the Lame's operator ${\cal L}_2[y]$ (end).
}
\begin{eqnarray*}
{\cal L}_2[y]&=&Q_4(x)y''+Q_3(x) y' +(\alpha\beta
x^2+\rho_1x+\rho_2)y= 0= (LD+M)(\bar L D+\bar M)y. \quad \bar
M=-\frac{3}{2} x^2+\bar B x+\bar C,\,\, M =
\frac{1}{2},\,\,\alpha=-\frac{3}{2},\,\,\beta=\frac{5}{2}.
\end{eqnarray*}
%}
\vspace{.5cm}
\begin{tabular}{||l|l|l|l|l|l||}
\hline \hline
 $L$ & ${\bar L}$ & $\bar B$ & $\bar C$ &$\rho_1$&$\rho_2$  \\
\hline\hline
 $x$&$(x-1)(x-a)(x-b)$&$a+b+1$&$-\frac{1}{2}(a+b+ab)$&$\frac{3}{2}(a+b+1)$&$-\frac{1}{4}(a+b+ab)$\\
 \hline
 $x-1$&$x(x-a)(x-b)$&$a+b$&$-\frac{1}{2}ab$&$\frac{3}{2}(2+a+b)$&$-\frac{1}{4}(4a+4b+ab)$\\
 \hline
 $x-a$&$x(x-1)(x-b)$&$b+1$&$-\frac{1}{2}b$&$\frac{3}{2}(2a+b+1)$&$-\frac{1}{4}(4a+b+4ab)$\\
 \hline
 $x-b$&$x(x-a)(x-1)$&$a+1$&$-\frac{1}{2}a$&$\frac{3}{2}(a+2b+1)$&$-\frac{1}{4}(a+4b+4ab)$\\
\hline \hline
\end{tabular}

\end{center}
} \label{tab03}
%\end{table}
 \end{sidewaystable}
\section{Factorization of the generalized Heun's equation ${\cal H}_2[y],
\, k=2$}
   The first extension of
   Heun's equation  with $k=2$ can be written as ($y(x)\equiv y$)
   \begin{equation}
   y''+\left[\frac{\gamma}{x}+\frac{\delta}{x-1}+
   \frac{\epsilon_{1}}{x-a}+\frac{\epsilon_{2}}{x-b}
   \right]y' +\frac{\alpha\beta x^{2}+\rho_{1}x+ \rho_{2}}{x(x-1)(x-a)(x-b)}y=0
   \end{equation}
   with the parameters linked by the Fuchs's condition $\alpha+\beta+1=
   \gamma+\delta+\epsilon_{1}+\epsilon_{2}$  or in polynomial way
   \begin{equation}\label{eq7}
   {\cal H}_{2}[y]=Q_{4}(x)y''+Q_{3}(x)y' +Q_{2}(x)y=0,
   \end{equation}
   where
   \begin{eqnarray}
   Q_{4}(x)&=&
   x(x-1)(x-a)(x-b)\nonumber\\&=&x^{4}-x^{3}(1+a+b)+x^{2}(a+b+ab)-x a b;\\
   Q_{3}(x)&=&\gamma(x-1)(x-a)(x-b)+\delta
   x(x-a)(x-b)\nonumber;\\
   &+&\epsilon_{1}x(x-1)(x-b) +\epsilon_{2} x(x-1)(x-a)\\
   Q_{2}(x)&=& \alpha\beta x^{2}+ \rho_{1}x+\rho_{2}.
   \end{eqnarray}
   The factorized form is
   %, like in
    %\cite{bi7,bi8},
    %with $D=\frac{\rm d}{{\rm d}x}$
    \begin{eqnarray}
    {\cal H}_{2}[y]&=&(LD+M)\left(\bar{L}D+\bar{M}\right)\cr
    &=& L\bar{L} D^{2}
    +\left(L\bar{L}'+L\bar{M}+ M\bar{L}\right)D
    +\left(L\bar{M}'+M\bar{M}\right).
     \end{eqnarray}

    % with the polynomials $L=L(x)$, $\bar{L}=\bar{L}(x)$, $M=M(x)$ and
    % $\bar{M}=\bar{M}(x)$ to be computed from (\ref{eq7}). The first relation
    % $L\bar{L}=Q_{4}(x)$ gives many possibilities for the decomposition of
    % $Q_{4}(x)$. Let us call $L_{r}\bar{L}_{4-r}$ the factor product; each $L_r$ of degree $r$
    % generates also many
     % choices involving  several singularities.

      Let us consider a peculiar simple situation
 \begin{equation}
       L=x(x-1),\qquad \bar{L}=(x-a)(x-b).
      \end{equation}
      The third relation in (\ref{eqn0}): $L\bar{M}'+M\bar{M}=Q_{2}(x)$ obliges $M$ and $\bar{M}$
      to be of degree $1$.
      Let $M=xA+B,$ $\bar{M}=x\bar{A}+ \bar{B}.$   The unknown $A$, $B$, $\bar{A}$, $\bar{B}$
      are given using also the second relation in (\ref{eqn0}).
      From
      $Q_{2}(x)=\bar{A}x(x-1)+(xA+B)(x\bar{A}+\bar{B})$, we get
      \begin{equation}
      \left\{
      \begin{array}{lll}
      \bar{A}+A\bar{A}&=&\alpha\beta,\\
       -\bar{A}+\bar{B}A + B\bar{A}&=&\rho_{1}.
        \end{array}
        \right.
      \end{equation}
      As in the Lam\'e's case, some kind of symmetry appears when
      permuting $L$ and $\bar L$ in the factorization, allowing to
      reduce by a factor $2$ the number of factorizations. Use of
      the adjoint operator ${\cal H}^\star$ of ${\cal H}$ is now
      useful. Indeed, the Lagrange adjoint ${\cal H}^\star$ of ${\cal H}$
      is given by \cite{bi1, bi18}
      \begin{eqnarray}
      {\cal H}^\star= Q^\star_{k+2} D^2 + Q^\star_{k+1} D + Q^\star_{k}
      \end{eqnarray}
 with
\begin{eqnarray}\label{eq41}
Q^\star_{k+2}= Q_{k+2},\;\;
 Q^\star_{k+1}= 2Q'^\star_{k+2}-Q_{k+1},\;\;
 Q^\star_{k}= {Q''}_{k+2} - Q'_{k+1} + Q_{k}
\end{eqnarray}
It is not very easy to write in general ${\cal H}^\star$ in the form
of the equation (\ref{eq13}) with star parameters $\alpha^\star,\;
\beta^\star, \; \rho^\star_i$, because the corresponding equations
are quadratic. But from the factorized form of ${\cal H}$,
computation of polynomials $M^\star$ and $\bar M^\star$ are trivial
from the relations:
\begin{eqnarray}
{\cal H}= (LD + M) (\bar LD+\bar M), \;\;
 {\cal H}^\star= (\bar
LD+\bar M)^\star (LD + M)^\star
\end{eqnarray}
and, from equation (\ref{eq41}), we get
\begin{eqnarray}
{\cal H}^\star&=& (-\bar LD+ \bar L'-\bar M) (-LD + L' - M)\cr &=&
(\bar LD+\bar M^\star) (LD + M^\star)
\end{eqnarray}
with
\begin{eqnarray}
 \bar M^\star=\bar M - \bar L',\; M^\star= M -  L'
\end{eqnarray}
generalizing the relation (\ref{eq17}).

 In tables 5 to 7, we give, in the $14$ situations,  the values of
$M,$ $\bar M,$ $\alpha,$ $\beta,$ $\rho_1$ and $\rho_2$ as well as
those corresponding  the other choices of $L$ and $\bar L.$ Of
course, the symmetry in the factorization of ${\cal H}_2$ is not so
rich as in the case of ${\cal L}_2$. Nevertheless, a quick sight in
the tables allows to classify families having the same
representation for $\rho_1$ and $\rho_2$ in terms of $A,\; \bar A,\;
B,\;\bar B$. See tables 5 - 7. The symmetry between $L$ and $\bar L$
can also be observed in all tables.

\begin{sidewaystable}[h]
%\begin{table}[htbp]
{\footnotesize
\begin{center}
\caption{Factorization of the Heun's operator ${\cal H}_2[y]$ }
\begin{eqnarray}
{\cal H}_2[y]&=&Q_4(x)y''+Q_3(x) y' +(\alpha\beta
x^2+\rho_1x+\rho_2)y= 0= (LD+M)(\bar L D+\bar M)y.\quad M=A x+B, \,
\, \bar M=\bar A x+\bar B.
\end{eqnarray}

%}
\vspace{.5cm}
\begin{tabular}{||l|l|l|l|l|l|l|l|l||}
\hline \hline
%& & & & & & & & & \\
 $L $& ${\bar L}$ & $A$ & ${\overline A}$ &   $B$ & ${\overline B}$ &$\alpha$ & $\beta$ \\
\hline\hline
%& & & & & & & & &\\
 $x(x-1)$&$(x-a)(x-b)$&
$\delta+\gamma$&$-2+\epsilon_1+\epsilon_2$&$-\gamma$&$a-\epsilon_2a+b-\epsilon_1b$&$-2+\epsilon_1+\epsilon_2$&$\gamma+\delta+1$\\
  \hline
$(x-a)(x-b)$&$x(x-1)$&$\epsilon_1+\epsilon_2$&$-2+\gamma+\delta$&$-\epsilon_2a-\epsilon_1b$&$1-\gamma$&
$\epsilon_1+\epsilon_2+1$&$-2+\gamma+\delta$\\
\hline
$x(x-a)$&$(x-1)(x-b)$&$\epsilon_1+\gamma$&$\delta-2+\epsilon_2$&$-\gamma
a$&$1-\epsilon_2-\delta
b+b$&$\delta-2+\epsilon_2$&$\gamma+\epsilon_1+1$\\
\hline
$(x-1)(x-b)$&$x(x-a)$&$\delta+\epsilon_2$&$-2+\epsilon_1+\gamma$&$-\delta
b-\epsilon_2$&$a-\gamma
a$&$1+\delta+\epsilon_2$&$-2+\epsilon_1+\gamma$\\
\hline
$x(x-b)$&$(x-1)(x-a)$&$\gamma+\epsilon_2$&$\delta-2+\epsilon_1$&$-\gamma
b$&$1-\epsilon_1-\delta
a+a$&$\delta-2+\epsilon_1$&$1+\gamma+\epsilon_2$\\
\hline
$(x-1)(x-a)$&$x(x-b)$&$\epsilon_1+\delta$&$\epsilon_2-2+\gamma$&$-\delta
a-\epsilon_1$&$b-\gamma
b$&$1+\epsilon_1+\delta$&$\epsilon_2-2+\gamma$\\
\hline\hline
\end{tabular}

\begin{tabular}{||l|l|l|l||}
\hline
 $L $& ${\bar L}$ &$\rho_1$&$\rho_2$\\
\hline\hline
 $x(x-1)$&$(x-a)(x-b)$&$-\bar A+B\bar A+A\bar B $&$B\bar B$\\
 \hline
$(x-a)(x-b)$&$x(x-1)$&$-(a+b)\bar A+B\bar
A+A\bar B$&$ab\bar A+B\bar B$\\
\hline $x(x-a)$&$(x-1)(x-b)$&$-a\bar A+B\bar
A+A\bar B$&$B\bar B$\\
\hline $(x-1)(x-b)$&$x(x-a)$&$-(1+b)\bar A+B\bar
A+A\bar B$&$b\bar A+B\bar B$\\
\hline $x(x-b)$&$(x-1)(x-a)$&$-b\bar A+B\bar
A+A\bar B$&$B\bar B$\\
\hline $(x-1)(x-a)$&$x(x-b)$&$-(1+a)\bar A+B \bar A
+A\bar B$&$a\bar A +B\bar B$\\
 \hline\hline
\end{tabular}
\end{center}
}\label{tab2}
%\end{table}
 \end{sidewaystable}

%%%%%%%%%%%%%%%%%%%%%%%%%%%%%%%%%%%%%%%%%%%%%%%%%%%%%%%%%%%%%%%%%%%%%%%%%%%%%%%%%%%%%%%%%%
%%%%%%%%%%%%%%%%%%%%%%%%%%%%%%%%%%%%%%%%%%%%%%%%%%%%%%%%%%%%%%%%%%%%%%%%%%%%%%%%%%%%%%%%%%

%\newpage
 \begin{sidewaystable}
%\begin{table}[htbp]
{\footnotesize
\begin{center}
\caption{Factorization of the Heun's operator ${\cal H}_2[y]$
(continuation). }
\begin{eqnarray}
{\cal H}_2[y]&=&Q_4(x)y''+Q_3(x) y' +(\alpha\beta
x^2+\rho_1x+\rho_2)y= 0= (LD+M)(\bar L D+\bar M)y. \quad M=A x^2+B
x+C,\,\,\bar M = \bar A.
\end{eqnarray}
%}
\vspace{.5cm}
\begin{tabular}{||l||l|l|l|l|l|l|l|l|l||}
\hline \hline
$ L$ & ${\bar L}$ & $A$ & $B$ &$C$&$\bar A$ &$\alpha$ & $\beta$ &$\rho_1$&$\rho_2$\\
\hline\hline
$(x-1)(x-a)(x-b)$&$x$&$\epsilon_1+\delta+\epsilon_2$&$-\delta
a-\delta b-\epsilon_2-\epsilon_2 a-\epsilon_1-\epsilon_1
b$&$\epsilon_2 a+\epsilon_1b+\delta
ab$&$-1+\gamma$&$\epsilon_1+\delta+\epsilon_2$&$-1+\gamma$&$B\bar A$&$C \bar A$\\
\hline
$x(x-a)(x-b)$&$x-1$&$\epsilon_1+\gamma+\epsilon_2$&$-\epsilon_2a-\gamma
a-\epsilon_1 b-\gamma b$&$\gamma a
b$&$-1+\delta$&$\epsilon_1+\gamma+\epsilon_2$&$-1+\delta$&$B\bar A$&$C\bar A$\\
\hline
$x(x-1)(x-b)$&$x-a$&$\delta+\gamma+\epsilon_2$&$-\epsilon_2-\gamma-\gamma
b-\delta b$&$\gamma
b$&$-1+\epsilon_1$&$\delta+\gamma+\epsilon_2$&$-1+\epsilon_1$&$B\bar A$&$C\bar A$\\
\hline $x(x-a)(x-1)$&$x-b$&$\epsilon_1+\delta+\gamma$&$-\gamma
a-\delta a-\gamma-\epsilon_1$&$\gamma
a$&$-1+\epsilon_2$&$\epsilon_1+\delta+\gamma$&$-1+\epsilon_2$&$B\bar A$&$C\bar A$\\
 \hline \hline
\end{tabular}
\end{center}
} \label{tab3}
%\end{table}
 \end{sidewaystable}

%%%%%%%%%%%%%%%%%%%%%%%%%%%%%%%%%%%%%%%%%%%%%%%%%%%%%%%%%%%%%%%%%%%
%%%%%%%%%%%%%%%%%%%%%%%%%%%%%%%%%%%%%%%%%%%%%%%%%%%%%%%%%%%%%%%%%%%
%\newpage
 \begin{sidewaystable}
%\begin{table}[htbp]
{\footnotesize
\begin{center}
\caption{Factorization of the Heun's operator ${\cal H}_2[y]$ (end).
}
\begin{eqnarray}
{\cal H}_2[y]&=&Q_4(x)y''+Q_3(x) y' +(\alpha\beta
x^2+\rho_1x+\rho_2)y= 0= (LD+M)(\bar L D+\bar M)y. \quad \bar M=\bar
A x^2+\bar B x+\bar C,\,\, M =  A.
\end{eqnarray}
%}
\vspace{.5cm}
\begin{tabular}{||l|l|l|l|l|l||}
\hline \hline
 $L$ & ${\bar L}$ & $A$ & $\bar A$ &$\bar B$&$\bar C$  \\
\hline\hline
 $x$&$(x-1)(x-a)(x-b)$&$\gamma$&$-3+\epsilon_1+\delta+\epsilon_2$&$2a+2b+2-\delta a-\delta b-\epsilon_2-\epsilon_2
 a-\epsilon_1-\epsilon_1b$&$-ab-a-b+\epsilon_2 a+\epsilon_1b+\delta
 ab$\\
 \hline
 $x-1$&$x(x-a)(x-b)$&$\delta$&$\gamma-3+\epsilon_1+\epsilon_2$&$2b-\epsilon_1b-\gamma b+2a-\gamma a-\epsilon_2
 a$&$-ab+\gamma a  b$\\
 \hline
 $x-a$&$x(x-1)(x-b)$&$\epsilon_1$&$-3+\delta+\gamma+\epsilon_2$&$2b-\delta b-\gamma
 b+2-\gamma-\epsilon_2$&$-b+\gamma
 b$\\
 \hline
 $x-b$&$x(x-a)(x-1)$&$\epsilon_2$&$-3+\epsilon_1+\gamma+\delta$&$2-\epsilon_1-\gamma+2a-\gamma a-\delta
 a$&$-a+\gamma a$\\
\hline \hline
\end{tabular}

\begin{tabular}{||l|l|l|l|l|l||}
\hline\hline
$L$ & ${\bar L}$ &$\alpha$ & $\beta$&$\rho_1$&$\rho_2$\\
\hline\hline
 $x$&$(x-1)(x-a)(x-b)$& $-3+\epsilon_1+\delta+\epsilon_2$&$2+\gamma$&  $\bar B+A\bar B$&$A\bar C$\\
\hline
 $x-1$&$x(x-a)(x-b)$& $\gamma-3+\epsilon_1+\epsilon_2$&$2+\delta$&   $-2\bar A+\bar B+A\bar B$&$-\bar B +A\bar C$\\
 \hline
 $x-a$&$x(x-1)(x-b)$& $-3+\delta+\gamma+\epsilon_2$&$2+\epsilon_1$&    $-2a\bar A+\bar B+A\bar B$&$-a\bar B+A\bar C$\\
 \hline
 $x-b$&$x(x-a)(x-1)$&  $-3+\epsilon_1+\gamma+\delta$&$2+\epsilon_2$&  $-2b\bar A+\bar B+A\bar B$&$-b\bar B+A\bar C$\\
\hline\hline
\end{tabular}

\end{center}
} \label{tab4}
%\end{table}
 \end{sidewaystable}

\section{Concluding remarks}
Some important features deserve to be pointed out.
\begin{enumerate}
\item In each case both solutions $y_i(x), \;(i=1, 2)$ of the
factorized equations are easily obtained solving first the equation
$$\bar L y'_i(x) + \bar M y_i(x) = 0.$$
In all Lam\'e's cases, these two solutions are given by
\begin{eqnarray}
y_1(x)=\sqrt{\bar L(x)},\;y_2(x) = \sqrt{\bar
L(x)}\int{{dx}\over{\bar L(x)\sqrt{L(x)\bar L(x)}}}.
\end{eqnarray}
We recover, of course, for $k= 1$, the $8$ "pseudo polynomials" of
Lam\'e \cite{petiau}
\begin{eqnarray}
y(x)= \sqrt{\bar L(x)} P_N(x), \; \mbox{(first type)},
\end{eqnarray}
 $P_N(x)$ being the corresponding Lam\'e's polynomials but now with
 appropriate parameters $\alpha,\,\beta$ and $\rho_i$.
 \item This approach does not give a new help in the search of a
 pecular solution of a generalized Heun's equation. But it gives a
 global approach of all solutions for arbitrary parameters
 ($\epsilon_i,\; \tau_i$).  This factorization defines in some way a
 parameter space splitted into many domains defined by "factor
 parameters" given in the several tables.
 \item The factorization becomes more and more difficult to handle
 with the increase of the number of $k+2$  singularities. Even with $k=
 3$, there are already $32$ different situations to investigate,
 what is irrelevant to be presented as it implies very long and complicated
analytical expressions for the polynomials $L,\;M,\;\bar L,\;\bar
M$. A Maple code is now under investigation for symbolic computation
for arbitrary number $k$ of singularities.
\end{enumerate}
\section*{Acknowledgements} The authors are thankful to Dr Alain
Moussiaux from the Facult\'es Universitaires Notre Dame de la Paix
(FUNDP), Namur, Belgium, for helpful discussions.


\begin{thebibliography}{00}
\addcontentsline{toc}{section}{References} \frenchspacing \small
\addtolength{\itemsep}{-4pt}
 \bibitem{bi1}
 E. L. Ince, Ordinary differential equation, Dover Publication Inc 1999
 \bibitem{bi2}
 T. Kimura, On Fuchsian differential equations reducible
 to hypergeometric equation by linear transformations, Funkcial. Ekvac. {\bf 13}
 (1970) 213-217.
 \bibitem{pool}E. G. C. Poole, Introduction of linear differential
 equations, Oxford University Press, Oxford Clarenton Press, 1936
 \bibitem{bi3}
 F. Bureau, Sur les equations differentielles lineaires du second
 ordre,  M\'emoires de la Soci\'et\'e Royale des Sciences de
 Li\`ege, $3^{ie}$ tome.
 \bibitem{bi4}
 G. H. Halphen, M\'emoire sur la reduction des equations differentielles lineaires aux
 formes integrales. Acad. Sc. Paris, $2^{ie}$ serie t {\bf $XXVIII$} 1984.
 \bibitem{bi5}
 M.  Picard, Trait\'e d'analyse T. III
 \bibitem{bi6}
 A. Ronveaux, Editor, Heun's Differential Equations,  Oxford University
 Press, Oxford and New York, 1995.
 \bibitem{bi7}
 A. Ronveaux, Factorisation  de l'op\'erateur de Heun,
 Rev. Questions Sci. {\bf 172} (2001) 409-416.
 \bibitem{bi8}
 A. Ronveaux, Factorization of the Heun's differential equation,
  Appl. Math. Comput. {\bf 141} (2003) 177-184.
 \bibitem{bi9}M. N. hounkonnou, A. Ronveaux and K. Sodoga,
 Factorization of some confluent Heun's differential equations, Appl. Math. Comput.
  {\bf 189} (2007) 816-820.
 \bibitem{bi10}
 F. M. Arscott, Heun's Equation, in Heun's Differential Equations
  (A. Ronveaux, Editor),   Oxford University Press, Oxford and New York, 1995,
  17-18 and 41-42
 \bibitem{bi11} R. S. Maier,
 Private communication (2002)
 \bibitem{bi12}
 M. Bronstein, An improved Algorithm for Factoring Linear Ordinary Differential
 Operators,
 Proceedings ISSAC'94, Oxford, UK, ACM Press, (1994) 336-340.
 \bibitem{bi13}
 F. Schwarz, A factorizzation Algorithm for Linear Ordinary Differential Equations,
 Proceedings of ISSAC'89, Oxford, ACM Press, (1989) 17-25.
 \bibitem{bi14}
 S. P. Tsarev, An algorithm for Complete Enumeration of All Factorization of a Linear
 Ordinary Differential Operator, Proceedings  of
 ISSAC'89, ACM Press, (1996) 176--189.
  \bibitem{bi15}
  M. Van Hoeij, Factorization of Differential Operators with Rational Functions
  Coeficients,  J. Symbolic Computation {\bf 24} (1997) 537-561.
\bibitem{bi15a} A. Pham Ngoc Dinh, Sur les fonctions
g\'en\'eralis\'ees de Lam\'e d'ordre $(n-2)$ associ\'ees \`a une
repr\'esentation induite de $E(n)$, Bull. Sc. Math. $2^e$, s\'erie
{\bf 104} (1980) 113-134.
  \bibitem{bi16} T. Guttmann,
 Private communication (1997).
 \bibitem{biphys} E. S. Cheb-Terrab,   New closed form solutions in terms of $_pF_q$ for families
of the General, Confluent and Bi-confluent Heun differential
equations,  J. Phys. A: Math.Gen. {\bf 37} (2004) 9923
 \bibitem{bi17} D. Basic and H. Schmid, Heun equation, Teukolsky equation and type-$D$
  metrics, J. Math. Phys. {\bf 48} (2007) 042502.
 \bibitem{bi18} G. Birkhoff and G-R. Rota, Ordinary differential
 equations, Ginn and Company (1962).
 \bibitem{petiau} G. Petiau, Les classes de solutions de Lam\'e,
 International Congress of Mathematics, Helsinki, (August 1978).
\end{thebibliography}
\end{document}